\documentclass[conference]{IEEEtran}

\usepackage{amsmath,amssymb,amsfonts}
\usepackage{graphicx}

\usepackage{cite}
\usepackage{bm}
\usepackage[hidelinks]{hyperref}
\usepackage{verbatim}

\expandafter\def\expandafter\normalsize\expandafter{%
    \normalsize%
    \setlength\abovedisplayskip{5pt}%
    \setlength\belowdisplayskip{5pt}%
}
\newcommand{\ignore}[1]{}

\begin{document}

\title{Scenario-Based Stochastic MPC for Energy Hubs with EV Fleets Under Persistent Grid Outages}

\author{
\IEEEauthorblockN{
Kobena Badu Enyam\textsuperscript{$\ast$}, 
Cara Koepele\textsuperscript{$\dagger$,$\ddagger$}, 
Timothy Asare\textsuperscript{$\dagger$}, 
Kevin Wallington\textsuperscript{$\dagger$}, 
John Lygeros\textsuperscript{$\dagger$}
}
\IEEEauthorblockA{
\textsuperscript{$\ast$}\textit{Ashesi University}, Accra, Ghana\\
\textsuperscript{$\dagger$}\textit{Automatic Control Laboratory, ETH Z\"{u}rich}, Z\"{u}rich, Switzerland\\
\textsuperscript{$\ddagger$}\textit{Urban Energy Systems Laboratory, EMPA}, Z\"{u}rich, Switzerland\\
kenyam@ashesi.edu.gh, \{ckoepele, tasare, kwallington, jlygeros\}@control.ee.ethz.ch
}
}

\maketitle

\begin{abstract}
Emissions reduction and resilience to outages motivate the adoption of renewable microgrids. Surprisingly, research integrating both probabilistic grid outages and electric vehicle (EV) charging requirements remains limited. This paper addresses this gap by developing a scenario-based stochastic model predictive controller (SMPC) for a microgrid energy hub comprising solar generation, battery storage, diesel backup, and an EV fleet connected to a weak grid. Grid outage and campus load scenarios are generated from a continuous-time Markov chain and a Gaussian Process, respectively. Using 2023 operational data from the Ashesi University Energy Hub in Ghana, we demonstrate that the SMPC achieves performance within 1\% of a perfect-forecast benchmark. In contrast, a naive MPC that assumes continuous grid availability offers no economic or sustainability advantage over rule-based control, with both incurring significantly higher costs and emissions than the SMPC. These results highlight that outage anticipation is essential for economic viability. Finally, we show that incorporating a deterministic buffer against EV consumption uncertainty eliminates over 90\% of state-of-charge violations with negligible impact on total operating costs.
\end{abstract}          


\section{Introduction}
Sites connected to weak electrical grids often rely on expensive, high-emission diesel generators during outages. In regions like Sub-Saharan Africa, adopting hybrid renewable-diesel systems is a promising strategy for improving reliability and reducing fuel costs~\cite{ayaburi2020measuring, olatomiwa2015economic}. Transitioning to electric vehicles (EV) further accelerates fuel reduction and fosters sustainability. However, the stochastic, high-power demands of EV fleets introduce operational challenges when the grid is inconsistent. Rule-based dispatch is commonly chosen to manage these systems due to their simplicity. However, these lack the foresight necessary to manage limited resources efficiently, leading to excessive diesel usage~\cite{restrepo2021optimization}. Optimization based energy management systems offer a systematic alternative, enabling the effective coordination of multiple energy carriers with renewables, storage, and backup generation under an optimal control framework~\cite{geidl2006ehub, parisio2014model}. 

Model Predictive Control (MPC) is commonly used for energy hub operation due to its ability to utilize forecasts and integrate costs and constraints ~\cite{smith2023nrgsys}. Deterministic MPC formulations are common for grid-connected systems~\cite{gan2021gpfor_detmpc, hu2021model}, but the inherent feedback from the receding horizon provides limited protection against incorrect predictions. Recent uncertainty-aware controllers utilize chance-constraints~\cite{babic2023nonparachancempc} or scenario-based stochastic MPC (SMPC)~\cite{micheli2023stochastic} to hedge against forecast errors. These approaches can be expanded to manage EV fleets, where variable charging profiles are addressed through distributionally robust~\cite{nguyen2023distributionally} or scenario-based~\cite{zandrazavi2022stochasticev} methods. However, these works typically assume a continuously available grid or fully islanded hub, ignoring settings that transition between connected and islanded states, e.g., during outages~\cite{arcos2024mpcecuadorislanded, hans2020riskaversempcislanded}. Only a few studies consider outages for grid-connected hubs. For instance,~\cite{tavakoli2018cvarresilience} uses manual outage scenarios, while~\cite{tobajas2022mpcresilience} employs scenario trees for non-critical load shedding. To the best of the authors' knowledge, existing literature has not yet integrated probabilistic outage modeling with the uncertain demands of an EV fleet. 

This paper presents a stochastic MPC framework for sustainably managing multi-source energy hubs connected to weak grids while serving an EV fleet, motivated by the sub-Saharan context. The framework explicitly accounts for probabilistic grid outages together with EV fleet energy requirements, enabling coordinated dispatch under uncertain grid availability. The approach is validated on a case study of the energy hub under development at Ashesi University in Ghana, using local operational data. Stochastic campus load is modeled using a Gaussian Process (GP), while grid outages are represented by a continuous-time Markov chain (CTMC). A robustness buffer is incorporated into the EV energy model to account for variability in trip energy requirements. Numerical results show that the proposed SMPC achieves operating cost performance close to a perfect-forecast benchmark and is robust to EV state of charge (SoC) plugin uncertainty. Additionally, diesel consumption is significantly reduced compared to both an outage unaware Naive MPC that assumes a continuously-available grid and a standard rule-based controller.

\section{Energy System Model}
\begin{figure}[!t]
  \centering
  \includegraphics[width=\columnwidth]{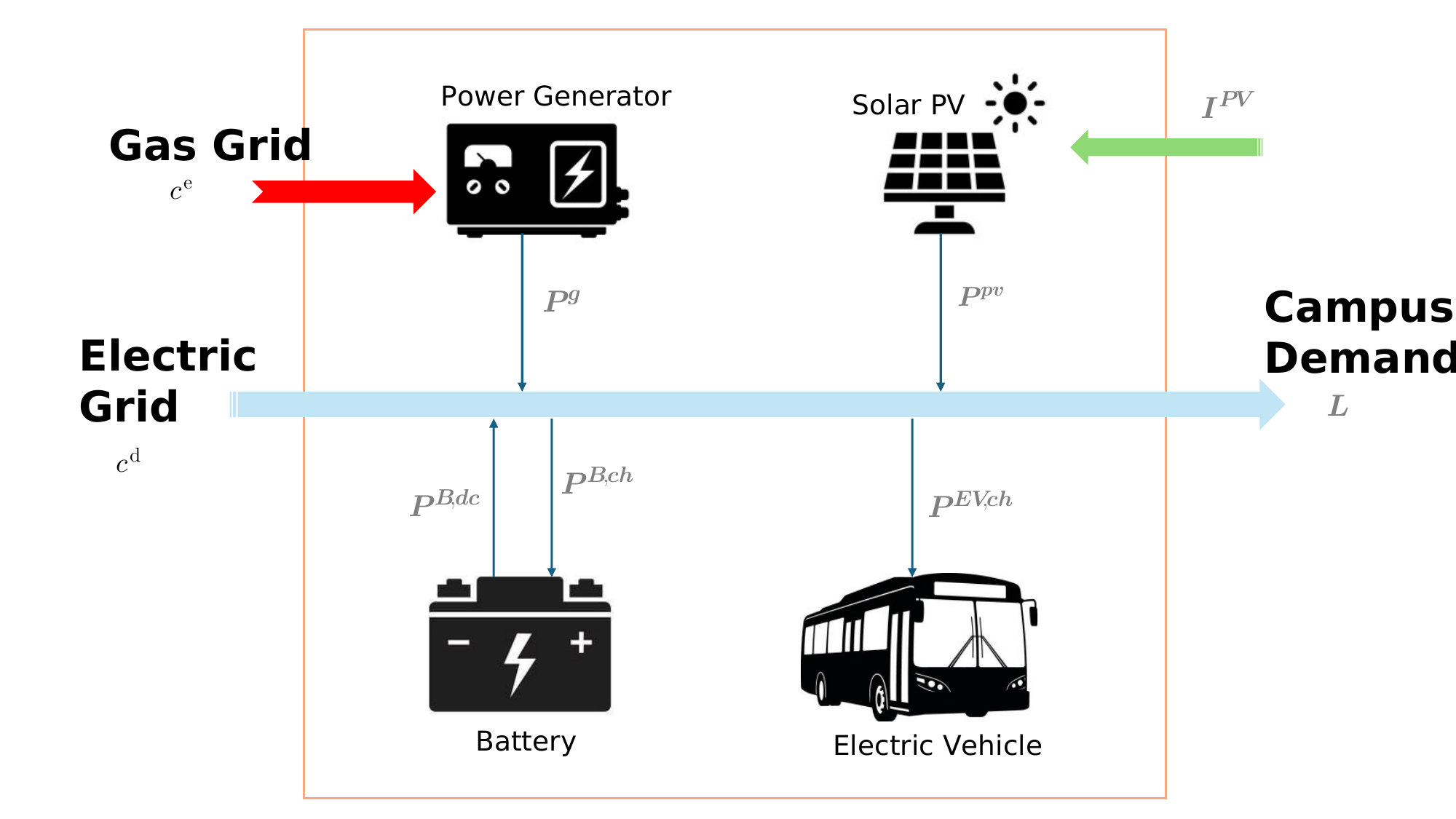}
  \caption{Energy-hub system model and power flows.}
  \label{fig:microgrid_model}
\end{figure}

We consider an energy hub equipped with photovoltaics (PV), a battery energy storage system (BESS), and backup diesel generators, as illustrated in Figure~\ref{fig:microgrid_model}. At each discrete timestep, $k$, at a sampling interval of $\Delta t$, the controller decides the dispatch of generation and storage technologies to supply a campus load and charge an EV bus fleet that serves the local public transport needs of the campus community. These decisions are subject to uncertain demands and grid availability. In addition to contending with unreliable grids, many countries in the sub-Saharan region prohibit grid injection. This regulatory constraint requires excess PV generation to be consumed, stored, or curtailed, and precludes vehicle-to-grid energy transfer.

\vspace{-0.01cm} 

The state of the system at timestep $k$ comprises the significant energy values of the system,
\begin{equation}
  \bm{x}_k =
  \begin{bmatrix}
    E_k^{\mathrm{B}}&
    (\bm{E}^{\mathrm{EV}}_k)^{\top}
  \end{bmatrix}^\top.
  \label{eq:state}
\end{equation}
This includes the stored energy in the battery, $E_k^{\mathrm{B}}$, and the vector of the energy states of $n$ buses, $\bm{E}_k^{\mathrm{EV}}=[E_k^{\mathrm{EV},i}]_{i=1}^{n}$. Throughout this paper, boldface symbols denote column vectors and $i$ indexes the vehicles in the EV fleet. 

The control input includes all power dispatch decisions,
\begingroup
\setlength{\arraycolsep}{1.5pt}
\begin{IEEEeqnarray}{rCl}
  \bm{u}_k =
  \begin{bmatrix}
    P_k^{\mathrm{e}} &
    P_k^{\mathrm{g}} &
    P_k^{\mathrm{PV}} &
    P_k^{\mathrm{B,ch}} &
    P_k^{\mathrm{B,dc}} &
    (\bm{P}_k^{\mathrm{EV}})^\top &
    L_k^{\mathrm{sh}}
  \end{bmatrix}^{\top}.
  \label{eq:input}
  \IEEEeqnarraynumspace
\end{IEEEeqnarray}
\endgroup
The components of $\bm{u}_k$ encompass grid imports $P_k^{\mathrm{e}}$, diesel generation $P_k^{\mathrm{g}}$, and PV injection after curtailment $P_k^{\mathrm{PV}}$. Storage operations are represented by the BESS charging and discharging powers, $P_k^{\mathrm{B, ch}}$ and $P_k^{\mathrm{B,dc}}$, while the vector $\bm{P}_k^{\mathrm{EV}}=[P_k^{\mathrm{EV},i}]_{i=1}^{n}$ captures individual EV charging rates. Emergency load shedding is denoted by $L_k^{\mathrm{sh}}$. 

The exogenous disturbances at time $k$ are
\begin{equation}
  \bm{w}_k =
  \begin{bmatrix}L_k & \theta_k & (\bm{z}_k^{\mathrm{EV}})^\top & (\bm{E}_k^{\mathrm{req}})^\top\end{bmatrix}^\top.
  \label{eq:exo}
\end{equation}
At timestep $k$, we define the campus load $L_k$, grid availability $\theta_k \in \{0,1\}$ (where $0$ denotes an outage), the vector $\bm{z}_k^{\mathrm{EV}}=[z_k^{\mathrm{EV},i}]_{i=1}^{n}$ whose $i^{\mathrm{th}}$ element is a binary variable indicating if vehicle $i$ is connected to a charger, and the vector $\bm{E}^{\mathrm{req}}_{k}=[E_k^{\mathrm{req},i}]_{i=1}^{n}$ which denotes each vehicle's energy consumption while driving. 

The BESS is modeled according to the dynamics and constraints: 
\begin{subequations}
\label{eq:ess-constraints}
\begin{gather}
E_{k+1}^{\mathrm{B}} = E_k^{\mathrm{B}} + \Delta t \left(\eta^{\mathrm{B}}_{\mathrm{\mathrm{ch}}} P_k^{\mathrm{B,ch}} - \frac{1}{\eta^{\mathrm{B}}_{\mathrm{dc}}} P_k^{\mathrm{B,dc}}\right), \label{eq:bess-dyn}\\
\underline{E}^{\mathrm{B}} \leq E_k^{\mathrm{B}} \leq \bar{E}^{\mathrm{B}}, \label{eq:bess-nrg}\\
0 \leq P_k^{\mathrm{B,ch}},\ P_k^{\mathrm{B,dc}} \leq  \bar{P}^{\mathrm{B}}. \label{eq:bess-pwr} 
\end{gather}
\end{subequations}
The state of charge is governed by charging and discharging powers, $(P_k^{\mathrm{B, ch}}, P_k^{\mathrm{B,dc}})$, and their associated efficiencies, $\eta^{\mathrm{B}}_{\mathrm{\mathrm{ch}}},\eta^{\mathrm{B}}_{\mathrm{dc}}\in(0,1)$. Operational limits are enforced with energy bounds, $(\underline{E}^{\mathrm{B}}, \bar{E}^{\mathrm{B}})$, and a power rating, $\bar{P}^{\mathrm{B}}$. We do not enforce a hard complementarity constraint between $P^{\mathrm{B, ch}}$ and $P^{\mathrm{B,dc}}$ because efficiency losses and degradation costs implicitly make simultaneous charge and discharge suboptimal. If required, adding binary complementarity constraints yields a Mixed Integer Quadratic Program. 
 
We adopt an individual model for the EV fleet. The energy storage dynamics and constraints for each EV, $i$, are
\begin{subequations}
\label{eq:ev_dyn}
\begin{align}
E^{\mathrm{EV},i}_{k+1} =\; &E^{\mathrm{EV},i}_k + \Delta t\, \eta^{\mathrm{EV}}_{\mathrm{ch}} P^{\mathrm{EV},i}_k -  E_{k}^{\mathrm{req},i}(1 - z^{\mathrm{EV},i}_k), \label{eq:ev_dyn_a} \\
    \underline{E}^{\mathrm{EV},i} - \;& \xi^i_k \leq E^{\mathrm{EV},i}_k \leq \bar{E}^{\mathrm{EV},i},\;0 \leq\; \xi^i_k \leq \underline{E}^{\mathrm{EV},i},
                      \label{eq:ev_dyn_b} \\
    0 \leq\;& P^{\mathrm{EV},i}_k \leq \bar{P}^{\mathrm{EV},i} z^{\mathrm{EV},i}_k.
                      \label{eq:ev_dyn_c}
\end{align}
\end{subequations}
The EV state of charge, $E^{\mathrm{EV},i}_k$, evolves according to the charging efficiency $\eta^{\mathrm{EV}}_{\mathrm{ch}}$ and the exogenous driving behaviors and is bounded by manufacturer-specified limits $[\underline{E}^{\mathrm{EV},i}, \bar{E}^{\mathrm{EV},i}]$. We introduce non-negative slack variable $\xi^i_k$ to permit soft violations of the lower energy bound.

The PV generation has a time-varying upper bound, 
\begin{equation}
  0\le P_k^{\mathrm{PV}}\le \bar{P}_k^{\mathrm{PV}}.
  \label{eq:pv_bound}
\end{equation}
The maximum PV generation, $\bar{P}_k^{\mathrm{PV}}$, is computed from the solar irradiance and ambient temperature, of which we assume perfect foresight.  

The power dispatch is subject to physical capacity limits for the utility connection and diesel generators:
\begin{equation}
0 \le P_k^{\mathrm{e}} \le \theta_k \bar{P}^{\mathrm{e}}, \quad 0 \le P_k^{\mathrm{g}} \le \bar{P}^{\mathrm{g}},
\label{eq:generation_bounds}
\end{equation}
where $\bar{P}^{\mathrm{e}}$ is rated transformer limit and $\bar{P}^{\mathrm{g}}$, the generator capacity. Grid imports are prohibited during outages with $\theta_k$.

Finally, the power balance of the energy hub is 
\begin{equation}
L_k = P_k^{\mathrm{e}} + P_k^{\mathrm{g}} + P_k^{\mathrm{PV}} + P_k^{\mathrm{B,dc}} - P_k^{\mathrm{B,ch}} - \sum_{i=1}^{n}P_k^{\mathrm{EV},i} + L_k^{\mathrm{sh}}.
\label{eq:power_balance}
\end{equation}

\section{Outage and Load Forecasting}

\subsection{Load Forecasting with a Gaussian Process} \label{sec:GP}

Campus electrical demand is forecast using a GP, $L(\phi) \sim \mathcal{GP}(\mu(\phi), \kappa(\phi, \phi'))$,
that is trained on historical data. The GP maps a feature vector $\phi_k$ to the campus load $L_k$, where
\begin{equation}
\phi_k =
\begin{bmatrix}
\sin\!\left(\frac{2\pi h_k}{24}\right),\ \cos\!\left(\frac{2\pi h_k}{24}\right),\ \sin\!\left(\frac{2\pi d_k}{365}\right),\\
\cos\!\left(\frac{2\pi d_k}{365}\right),\ b_k
\end{bmatrix},
\label{eq:time_features}
\end{equation}
uses the hour-of-day $h_k$, day-of-year $d_k$, and a boolean $b_k=\{0,1\}$ to denote a weekday or weekend/holiday respectively from the timestep $k$. A composite Exp-Sine-Squared plus White noise kernel, $\kappa(\cdot, \cdot)$, captures periodicity and measurement noise. The GP is updated monthly via a three-month sliding window: it is trained on the second and third preceding months, with hyperparameters tuned by maximizing the log-marginal likelihood on the most recent month. At each $k$, point estimates $\hat{L}_k \sim \mathcal{N}(\mu_k^L, \sigma_k^L)$ are sampled from the posterior and clipped to non-negative values. Since $\phi$ is deterministic, independent samples, $s$, construct load trajectory scenarios $L^{(s)}_{0:N}$ over the prediction horizon $N$.

\subsection{Outage Forecasting with a Continuous-Time Markov Chain}
Grid availability $\theta_k\in\{0,1\}$ is modeled as a two-state continuous-time Markov chain with outage rate $\psi$ (ON$\rightarrow$OFF) and restoration rate $\nu$ (OFF$\rightarrow$ON). This is a standard method of outage modeling in grid reliability studies such as~\cite {modarres2016reliability}. The corresponding discrete-time transition probabilities are 
\begin{equation}
p_{10}=1-e^{-\psi\Delta t},\qquad p_{01}=1-e^{-\nu\Delta t}.
\label{eq:ctmc_transitions}
\end{equation}
Each grid availability scenario $\theta_{0:N}^{(s)}$ is simulated forward over the prediction horizon, $N$, using the transition probabilities, initialized at the measured current state to capture frequency and severity of outages as obtained from historical data. 

\subsection{EV Energy Requirement Model}
\label{subsec:ev_model}
The per-timestep energy requirement $E_{k}^{\mathrm{req},i}$ in~\eqref{eq:ev_dyn}
captures the expected energy depletion of vehicle $i$ during trips.
We model this as
\begin{equation}
    E_{k}^{\mathrm{req},i} = (1 + \alpha)\, P^{\mathrm{loss},i}_{\mathrm{avg}} \,\Delta t,     P^{\mathrm{loss},i}_{\mathrm{avg}} =
    \frac{\sum_{j \in \mathcal{T}_i} E^{\mathrm{trip},i}_{j}}
         {\sum_{j \in \mathcal{T}_i} T^{\mathrm{trip},i}_{j}},
    \label{eq:ereq}
\end{equation}
where $P^{\mathrm{loss},i}_{\mathrm{avg}}$ is the route-averaged power draw over a set of historical trips, $\mathcal{T}_i$, and $\alpha
\geq 0$ is a conservativeness factor. The energy consumed on trip $j$ of route $i$ is $E^{\mathrm{trip},i}_{j}$, and the corresponding trip duration in hours is $T^{\mathrm{trip},i}_{j}$. 

To account for consumption variability driven by congestion, weather, and driver behavior, the parameter $\alpha$ scales up the nominal depletion rate. Since EVs are not yet operational at the Ashesi site, $P^{\mathrm{loss},i}_{\mathrm{avg}}$ is derived from existing diesel vehicle records, a conservative proxy as diesel buses are generally less efficient. Lack of available data prevents the estimation of per-vehicle consumption distributions, so a probabilistic bound would require unverifiable assumptions. The deterministic scaling in~\eqref{eq:ereq} instead provides a tunable margin whose impact is analyzed via a parametric sweep in subsection~\ref{subsec:microgrid_sim}. While $\alpha$ is ideally calibrated to each vehicle's historical coefficient of variation, this study uses a uniform $\alpha$ across the fleet.

\section{Energy Hub Controller Formulations}
We introduce three controller formulations, reflecting the current industry standard via a rule-based controller (RBC), a deterministic MPC, and our proposed uncertainty aware scenario-based MPC. 
\subsection{Rule-Based Control}
The RBC serves load demands and EV charging with the following resources in descending priority: PV, BESS, grid, and diesel. The BESS is only charged with excess PV after fulfilling the loads. EV charging is modeled as a flexible load, in the sense that the $\alpha$-conservative EV requirement can be arbitrarily distributed during timesteps when the EV is plugged in, subject to charger and capacity constraints. The RBC attempts to provide the maximum charge power with all resources except for diesel until the energy requirement is fulfilled. If the available power is less than total maximum charge power over all plugged-in EVs, the RBC prioritizes charging vehicles first by the earliest departure and then by the highest remaining energy requirement. Diesel consumption for EV charging is strictly reserved as a contingency in the final timestep before departure.

\subsection{Deterministic MPC}
MPC dictates the power dispatch for the energy by solving a finite-horizon optimization problem that aims to minimize energy purchases as well as load shedding or shortfall, BESS degradation, and PV curtailment, subject to operational constraints and disturbance forecasts. Let $p=\{\bm{x}, \bm{u}, \bm{\xi}\}$ collect the decision variables over a prediction horizon, $\mathcal{K}=\{0,\ldots,N-1\}$, and $\mathcal{P}$ be the full constraint set for a given deterministic disturbance forecast, $\omega$, as follows:
\begin{equation*}
    \mathcal{P}(\omega) :=\ \{p \; | \; \eqref{eq:ess-constraints}-\eqref{eq:power_balance},\eqref{eq:ereq}, \text{ hold } \forall k\in\mathcal{K} \text{ given }\omega\}.
\end{equation*}

At each timestep, the following optimization problem is solved, and only the first action is implemented:
\begin{align}
\begin{split}
\min_{\{x,u,\xi\}}
\;&
\overbrace{\sum_{k=0}^{N-1} l_k + l_k^{\mathrm{EV}}+ l^{\mathrm{term}}}^{J^N_t}  \text{s.t.}\; p\in\mathcal{P}(\omega),\ x_0 = x_t,
\end{split}
\label{eq:dmpc_problem}
\end{align}
Here, $l_k$ and $l^{\mathrm{term}}$ are the stage cost and terminal cost of the stationary energy system, respectively, and $l_{\mathrm{EV}}$ is the EV fleet cost. The initial measured state is $x_t$, and $J^N_t$ is the overall MPC objective at timestep $t$ for a horizon of $N$. The optimization problem \eqref{eq:dmpc_problem} ensures feasibility by including load shedding and EV shortfall slack variables, which are heavily penalized to discourage their use under normal operating conditions. The per-stage objective is
\begin{equation}
  \begin{aligned}
  l_k = \Bigl(&c^{\mathrm{e}} P_k^{\mathrm{e}} + \tfrac{c^{\mathrm{d}}}{\eta_{\mathrm{g}}} P_k^{\mathrm{g}} + \rho_{\mathrm{deg}} P_k^{\mathrm{B,dc}} + \\ &\rho_{\mathrm{sh}} L_k^{\mathrm{sh}} +\rho_{\mathrm{curt}} (\bar{P}_k^{\mathrm{PV}}-P_k^{\mathrm{PV}})\Bigr)\Delta t
  \end{aligned}
  \label{eq:stage_cost}
\end{equation}
where $c^{\mathrm{e}}$ is the electricity price, $c^{\mathrm{d}}$ is the diesel price, $\eta_{\mathrm{g}}$ is the average generator efficiency. Nonnegative weights $\rho_{\mathrm{deg}}$, $\rho_{\mathrm{curt}}$, and $\rho_{\mathrm{sh}}$ penalize battery cycling, PV curtailment, and load shedding, respectively. All prices are assumed to be known. The terminal cost, $l^{\mathrm{term}}$, quadratically penalizes deviation of the terminal BESS energy from a desired reference $E_{\mathrm{ref}}^B$:
\begin{equation}
  l^{\mathrm{term}} := \max(\left(E_{\mathrm{ref}}^B -E_N^{\mathrm{B}}\right),0)^2 \rho_{\mathrm{term}} ,
  \label{eq:terminal_cost}
\end{equation}
with weight $\rho_{\mathrm{term}}\ge 0$. The reference energy is typically set to the initial BESS energy, $E_0^{\mathrm{B}}$, to discourage depleting the battery at the end of the horizon.

The EV cost contains a shortfall penalty and a regularization cost for each bus:
\begin{equation}
  l_k^{\mathrm{EV}}= \sum_{i=1}^n \rho^{\mathrm{EV}}_{\mathrm{viol}} \xi_k^i + l^{\mathrm{EV},i}_{k,\mathrm{reg}}.
  \label{eq:ev_shortfall_cost}
\end{equation}
The weight $\rho^{\mathrm{EV}}_{\mathrm{viol}}\geq 0$ represents the price of violating the minimum EV energy bound. The regularization term, $l_{k,\mathrm{reg}}^{\mathrm{EV},i}$, is introduced to discourage aggressive EV charging and excessive step-to-step variation in charging decisions under tight operating constraints such as outages. Specifically, we penalize both charging-power increments and magnitude quadratically motivated by a similar approach in~\cite{houbbadi2019quadratic}:
\begin{align}
\begin{split}
  l_{k,\mathrm{reg}}^{\mathrm{EV},i} = &\lambda_{\Delta P}^{\mathrm{EV}}\left(P_k^{\mathrm{EV},i}-P_{k-1}^{\mathrm{EV},i}\right)^2+\lambda_{\mathrm{pk}}^{\mathrm{EV}}\left(P_k^{\mathrm{EV},i}\right)^2,
\end{split}
\label{eq:ev_reg}
\end{align}
Here, $\lambda_{\Delta P}^{\mathrm{EV}}\ge 0$ penalizes rapid changes in EV charging power between consecutive timesteps to smooth out the charging profile, while $\lambda_{\mathrm{pk}}^{\mathrm{EV}}\ge 0$ penalizes high instantaneous charging powers to limit charging peaks. This promotes an even EV charging profile across the prediction horizon.

At each timestep, the current state is measured and the disturbance forecasts are updated in optimization problem \eqref{eq:dmpc_problem}, which is then solved and the first action is applied. 

\subsection{Scenario-Based Stochastic MPC} 
Stochastic optimization incorporates the probabilistic characteristics of disturbances in the objective and constraints to facilitate uncertainty-aware decision-making. While analytical formulations provide exact solutions, they are typically limited to specific distribution classes, such as Gaussian noise, and scale poorly under multiple, heterogeneous sources of uncertainty. To integrate our GP campus load model with the CTMC outage model, we adopt a scenario-based framework. This approach utilizes Monte Carlo sampling to represent complex, generic disturbances through a finite set of trajectories, which emulates the behavior of each stochastic process while remaining computationally tractable.

At each real time $t$, we generate $M$ scenarios
\begin{equation}
  \omega^{(s)} := \{L_k^{(s)},\theta_k^{(s)}\}_{k=0}^{N-1},\qquad s\in\mathcal{M}=\{1,\ldots,M\},
  \label{eq:omega_s}
\end{equation}
using the GP load model and CTMC outage model for the set of scenarios, $\mathcal{M}$. Over the horizon, the disturbances $z_k^{\mathrm{EV},i}$ and $E_k^{\mathrm{req},i}$ are treated as deterministic, derived from the timetable and the buffered energy requirement in \eqref{eq:ereq}.

For each scenario, we optimize trajectories $p_k^{(s)}=\{\bm{x}_k^{(s)},\bm{u}_k^{(s)},\bm{\xi}_k^{(s)}\}$ satisfying $\mathcal{P}(\omega^{(s)})$. The initial states for all scenarios are constrained to be the measured state, $x_t$, at the current timestep, $t$. To ensure the controller is readily implementable, we additionally constrain the decision variables to be identical for the first step across all scenarios such that only one optimal action, $u_0$, to apply next is determined:
\begin{equation}
  x^{(s)}_0 = x_t,\; u_0^{(s)} = u_0,\qquad s\in\mathcal{M},
  \label{eq:nonanticipativity}
\end{equation}
The subsequent actions $u_k^{(s)}$ for $k\ge 1$ may vary across scenarios, optimistically allowing the decisions at time $k$ to depend on the realization of the disturbance at future timesteps. We adopt this formulation as it greatly simplifies the problem complexity compared to non-anticipative alternatives such as policies or scenario trees. 

The scenario-based optimization problem for time $t$ is:
{\small
\begin{equation}
\label{eq:smpc_problem}
\min_{\{x,u,\xi\}}\ \frac{1}{M} \sum_{s=1}^{M} J^{N,(s)}_t\;
\text{s.t.}\ p^{(s)}\in\mathcal{P}(\omega^{(s)}), \eqref{eq:nonanticipativity}, \forall s\in\mathcal{M},
\end{equation}
}
Here, $J_t^{N,(s)}$ represents the operation cost for scenario $s$, with the average of this cost over all scenarios as the objective. While this study does not enforce satisfaction requirements on load shedding or EV charging shortfalls, probabilistic guarantees on feasibility can be formally provided with chance constraints, which are often paired with scenario-based objectives. All constraints are affine, and the cost is a sum of linear terms and quadratic penalties, rendering \eqref{eq:smpc_problem} a convex quadratic program. Similar to the deterministic setting, the controller applies only the optimal shared first-step action $u_0^\star$ and repeats \eqref{eq:smpc_problem} at the next timestep.

\section{Numerical Simulation and Results}
\subsection{Simulation Setup and Evaluation Protocol}
A full-year hourly dataset from the Ashesi University energy hub (January–December 2023) is used to generate forecasts and evaluate controller performance. The dataset comprises campus electrical demand, weather data used to derive available PV power, and local vehicle schedules. The raw timeseries data is temporally aligned and resampled to a $1~$h interval. This data is then used to train and validate the GP for use on a monthly basis as described in Section \ref{sec:GP}. Grid outage parameters for the CTMC model are approximated from resident surveys on outage frequency and duration in 2023. 

The simulation assumes that existing diesel buses are replaced with EVs. The nominal EV availability trajectories and required energies, used by the controllers, are derived from the existing vehicle schedules, from which the average depleted powers during drives are computed.The performance of the controller is evaluated with respect to a realized schedule. The realized arrival times, dwell durations, and charging demands for each charging session are randomized via discrete and continuous Gaussian models centered on the nominal schedule. 
\begin{table}[t]
\centering
\footnotesize
\caption{Objective Parameter Values in GHS/kWh}
\label{tab:cost_params}
\renewcommand{\arraystretch}{1.5}
\setlength{\tabcolsep}{4pt}
\begin{tabular}{|r*{8}{c}|}
\hline
\textbf{Parameter} &$c^{\mathrm{e}}$ & $c^{\mathrm{d}}$ & $\rho_{\mathrm{sh}}$ & $\rho_{\mathrm{curt}}$ & $\rho_{\mathrm{deg}}$ & $\rho_{\mathrm{viol}}^{\mathrm{EV}}$ &  $\lambda_{\Delta P}^{\mathrm{EV}\text{a}}$ & $\lambda_{\text{pk}}^{\mathrm{EV}\text{a}}$ \\
\hline
\textbf{Value} & $2.5$ & $4.12$ & $1000$ & $0.05$ & $0.02$ & $50$ & $5.0$ & $0.02$ \\
\hline
\multicolumn{9}{l}{$^{\text{a}}$ parameter is unitless.}
\end{tabular}
\end{table}
The proposed SMPC is compared against a perfect-forecast MPC (PF-MPC), a Naive MPC (NMPC) and the industry-standard RBC. All MPC formulations have a prediction horizon of $N=24$~h and a sampling time of $\Delta t=1$~h. PF-MPC solves the optimization problem \eqref{eq:dmpc_problem} at every timestep with knowledge of the true future values of the campus load and outage disturbances, representing an unattainable lower bound on the energy hub costs under our MPC controllers. NMPC is a variant of the SMPC that retains the same optimization structure but assumes uninterrupted grid availability over the prediction horizon. All MPC controllers follow the EV energy requirement model presented in \eqref{eq:ereq}. To characterize the trade-off between satisfying EV demand and operating cost, $\alpha$ is varied from $0$ to $0.4$.

All cost parameters are summarized in Table \ref{tab:cost_params}. Parameter $c^{\mathrm{e}}$ is the electricity price and $c^{\mathrm{d}}$ is the diesel price, both based on 2025 data from Ghana \cite{PURC2025,NPA2025}. All other parameters are determined heuristically. SMPC and NMPC use $M=30$ disturbance scenarios, each comprising a GP-sampled load trajectory and a CTMC-generated grid-availability trajectory over the same horizon. 

The controllers are compared using microgrid operating costs, primary $\mathrm{CO}_2$ emissions from diesel consumption, and EV violation rates. For economic evaluation, we use the optimization objective without the EV and terminal costs, i.e., $J^{S,(s)}_t-l^{\mathrm{EV},(s)} - l^{\mathrm{term}}$, on the closed-loop response to the ``true" scenario comprising the test dataset and an outage scenario drawn from the assumed CTMC model. The simulation horizon, $S$, is one month long. $\mathrm{CO}_2$ emissions are computed from realized diesel dispatch using:
\begin{equation}
    \mathrm{CO}_2 = F_{\mathrm{g}} \Delta t \sum_{t} P_t^{\mathrm{g}},
\end{equation}
where $F_{\mathrm{g}} = 0.8 \, \text{kg CO}_2/\text{kWh}$ as indicated in~\cite{seforall2021carbon}. EV service violations are quantified by the lower-bound violation rate, which is the total number of violations divided by the total number of timesteps over the simulation horizon. Four representative months were selected to analyze the controller behavior across distinct operating regimes of campus demand and PV availability. January and July represent the periods of highest and lowest average PV generation, respectively, while April and June correspond to the highest and lowest average campus load, respectively.

\subsection{Microgrid Simulations}
\label{subsec:microgrid_sim}
\begin{figure}[!t]
  \centering
  \includegraphics[width=\columnwidth]{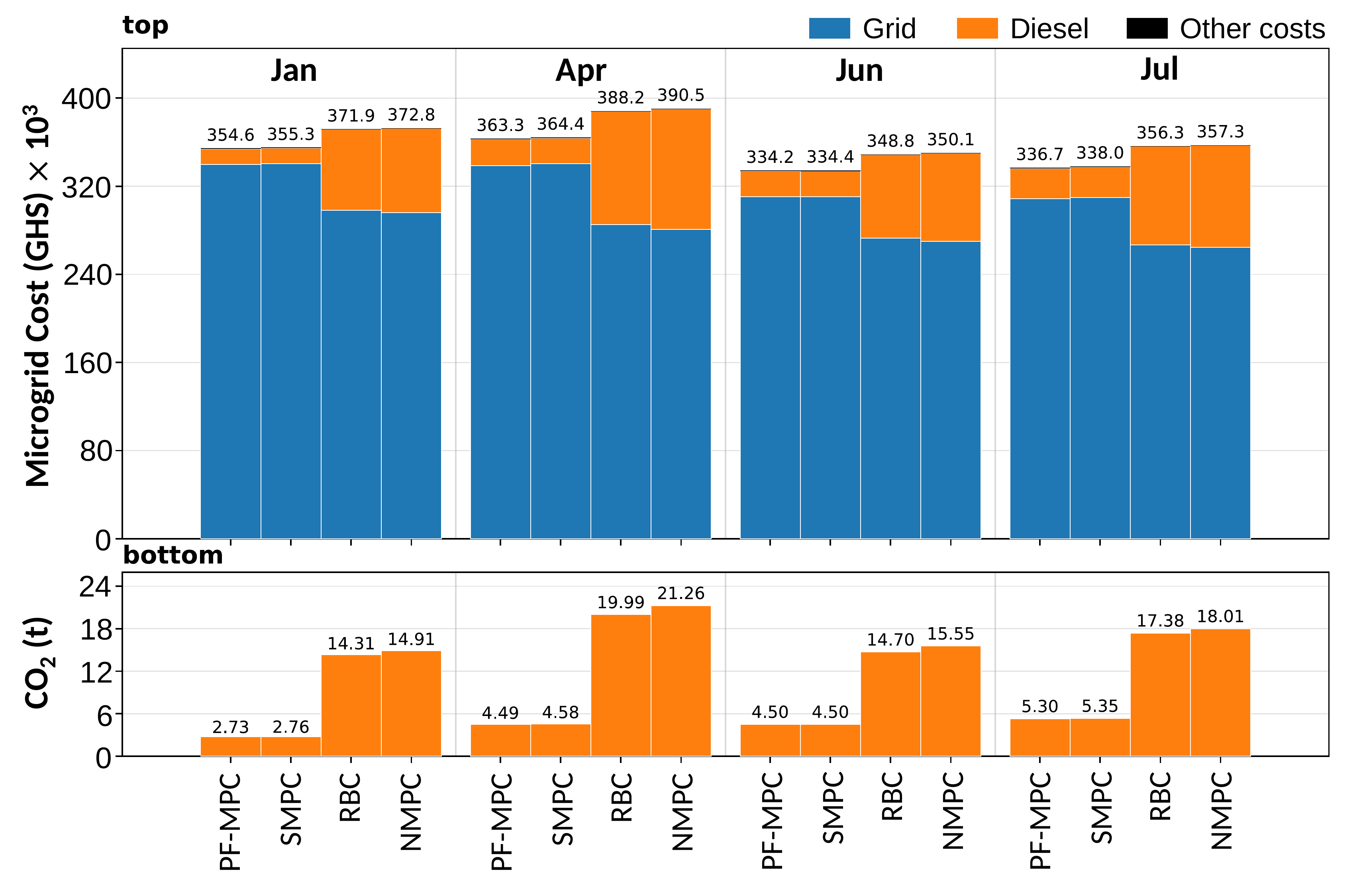}
  \caption{(top) Operating-cost comparison (Other costs = battery degradation + PV curtailment + load shedding) and (bottom) diesel-related CO$_2$ emissions ($\alpha = 0.4$.)}
  \label{fig:microgrid_comp}
\end{figure}

\begin{figure}[!t]
  \centering
  \includegraphics[width=\columnwidth]{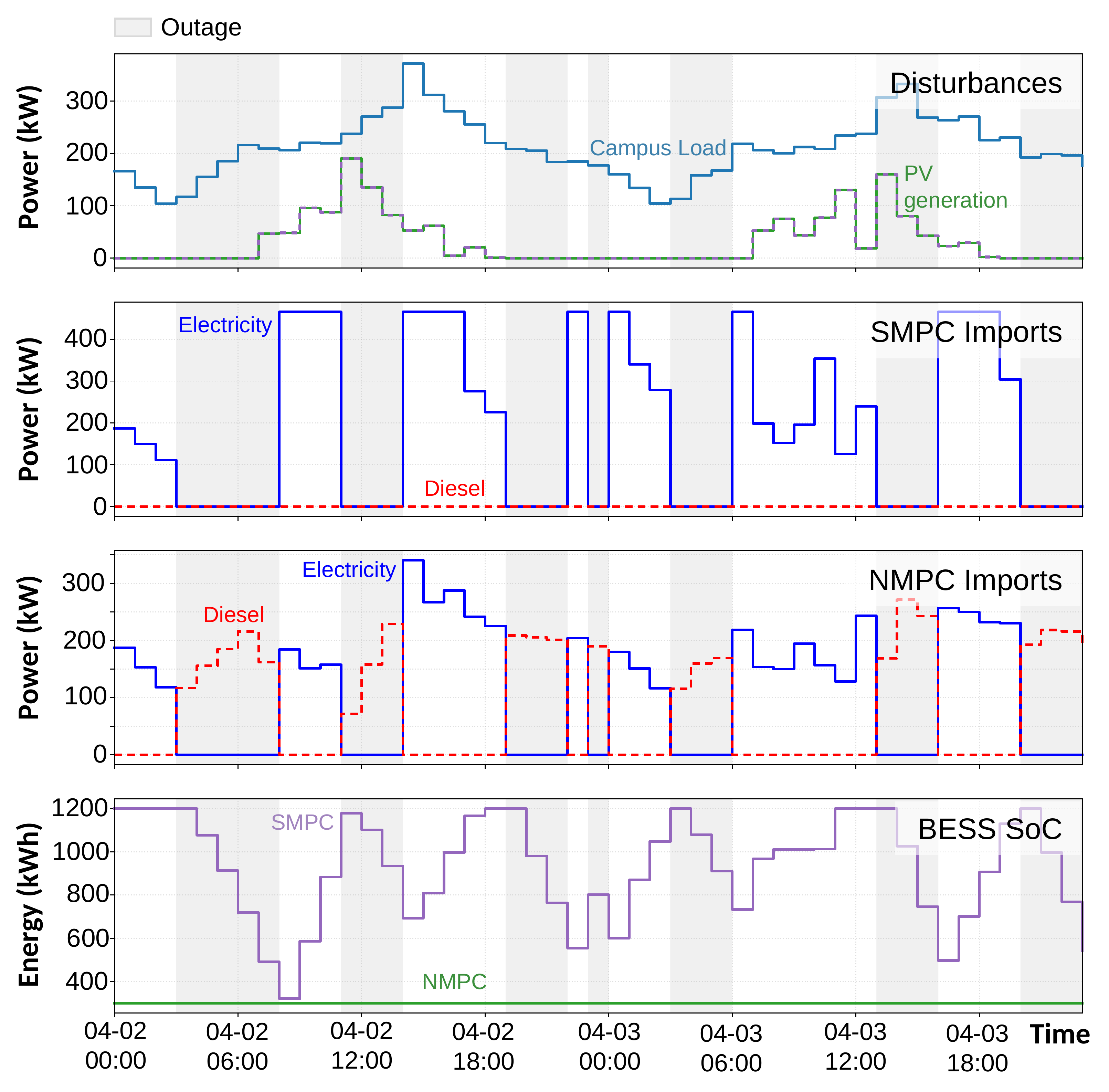}
  \caption{48-h representative window dispatch with $\alpha=0.4$.}
  \label{fig:window_48h}
\end{figure}

\hyperref[fig:microgrid_comp]{Figure~2} compares controller performance. The SMPC closely matches the PF-MPC benchmark, staying within 0.39\% of its cost and 2\% of its emissions. In contrast, the RBC and NMPC perform substantially worse, with cost increases exceeding 6.85\% and diesel-related CO$_2$ emissions increasing by over 420\%, relative to the PF-MPC. Since the only distinction between the SMPC and the NMPC is the inclusion of the outage model, these results demonstrate that outage-awareness is essential for sustainable and economic energy dispatch. Furthermore, the SMPC emission reductions are achieved with the existing local price structures, indicating that supplemental emissions penalties are unnecessary to discourage diesel usage.

\hyperref[fig:window_48h]{Figure~3} illustrates the operation of the SMPC versus the NMPC over a 48-hour window. While the SMPC utilizes the BESS during grid outages, it does not fully charge the BESS at every opportunity when the grid is available. Instead, SMPC leverages the probabilistic model to hedge charging levels against anticipated risk. In contrast, the NMPC perceives the grid as a constant, low-cost source. Consequently, the NMPC finds no incentive to store energy except in the case of surplus PV generation, leaving it reliant on diesel during grid failures. These behaviors show that effective BESS utilization is necessary in this context, which is enabled by outage consideration. 
\begin{figure}[!t]
  \centering
  \includegraphics[width=0.9\columnwidth]{Figure4_edit.pdf}
  \caption{Pareto of microgrid operating costs and EV violations due to varying the EV-charging buffer, $\alpha$, in April 2023 for: (top) all controllers and (bottom) SMPC only.}
  \label{fig:pareto}
\end{figure}

\hyperref[fig:pareto]{Figure~4} displays the Pareto trade-off between operating costs and EV service reliability in April 2023. While all controllers avoid EV battery depletion due to the manufacturer lower bound, the SMPC and NMPC significantly outperform the RBC in terms of service reliability. The RBC incurs violations nearly 9\% higher than the PF-MPC benchmark, whereas the SMPC and NMPC stay within 1.11\% and 1.48\%, respectively. Notably, though the NMPC has slightly higher cost and emissions compared to RBC, the NMPC provides superior requirement satisfaction, highlighting the benefit of encoded optimization constraints over simple rules. The buffer $\alpha$ has an efficient impact: increasing its value from 0 to 0.4 eliminates 92\% of SMPC violations at a cost increase of only 0.05\%. However, a sharp increase in the Pareto slope beyond $\alpha=0.3$ suggests diminishing returns for higher buffers. 

\section{Conclusion}
This paper presents a scenario-based stochastic MPC framework for energy hubs connected to weak grids that serve EV fleets. Grid outages are modeled with a continuous-time Markov chain and campus loads are forecast with a Gaussian process for generating disturbance scenarios, while a deterministic buffer hedges against EV consumption variability. Validated on a full year of operational data from the Ashesi University energy hub, the proposed SMPC operates within $0.39\%$ of a perfect-forecast benchmark in cost. A central finding is that without outage consideration, optimization alone provides no meaningful advantage: the Naive MPC, which assumes continuous grid availability, incurs up to $6.85\%$ higher cost and $446\%$ more diesel-related CO\textsubscript{2} emissions than the PF-MPC, comparable to the industry-standard RBC. Additionally, a simple deterministic buffer on EV energy requirements reduces EV SoC violations by over $90\%$ at negligible cost. These results demonstrate that outage-aware dispatch is essential for the economic and sustainable integration of EV fleets in unstable grid environments. Future work will incorporate PV forecast uncertainty, battery degradation models and real-time telemetry for field deployment.

\section*{Acknowledgment}
This work was supported as a part of NCCR Automation, a National Centre of Competence in Research, funded by the Swiss National Science Foundation (grant number 51NF40\_225155) and ETH for Development. We thank Lily A. Afriyie and Afari A. Antwi for providing data on the Ashesi bus fleet. Gemini 3 Flash (Google) was used to suggest alternative phrasing for text clarity, which the authors reviewed and modified for the final version of the paper.

\bibliography{references}

@article{parisio2014model,
  title        = {A model predictive control approach to microgrid operation optimization},
  author       = {Parisio, Alessandra and Rikos, Evangelos and Glielmo, Luigi},
  journal      = {IEEE Transactions on Control Systems Technology},
  volume       = {22},
  number       = {5},
  pages        = {1813--1827},
  year         = {2014},
  publisher    = {IEEE}
}

@article{ayaburi2020measuring,
  title        = {Measuring ``Reasonably Reliable'' access to electricity services},
  author       = {Ayaburi, John and Bazilian, Morgan and Kincer, Jacob and Moss, Todd},
  journal      = {The Electricity Journal},
  volume       = {33},
  number       = {7},
  pages        = {106828},
  year         = {2020},
  publisher    = {Elsevier}
}

@article{olatomiwa2015economic,
  title        = {Economic evaluation of hybrid energy systems for rural electrification in six geo-political zones of Nigeria},
  author       = {Olatomiwa, Lanre and Mekhilef, Saad and Huda, ASN and Ohunakin, Olayinka S},
  journal      = {Renewable Energy},
  volume       = {83},
  pages        = {435--446},
  year         = {2015},
  publisher    = {Elsevier}
}

@article{restrepo2021optimization,
  title        = {Optimization-and rule-based energy management systems at the Canadian renewable energy laboratory microgrid facility},
  author       = {Restrepo, Mauricio and Ca{\~n}izares, Claudio A and Simpson-Porco, John W and Su, Peter and Taruc, John},
  journal      = {Applied Energy},
  volume       = {290},
  pages        = {116760},
  year         = {2021},
  publisher    = {Elsevier}
}

@article{geidl2006ehub,
  title={Operational and structural optimization of multi-carrier energy systems},
  author={Geidl, Martin and Andersson, G\"oran},
  journal={European Transactions on Electrical Power},
  volume={16},
  pages={463--477},
  year={2006},
  publisher={Wiley InterScience}
}

@article{smith2023nrgsys,
  title={Control of Multicarrier Energy Systems from Buildings to Networks},
  author={Smith, Roy S. and Behrunani, Varsha and Lygeros, John},
  journal={Annual Review of Control, Robotics, and Autonomous Systems},
  volume={6},
  pages={391--414},
  year={2023},
  publisher={Annual Reviews}}

@article{hu2021model,
  title={Model predictive control of microgrids--An overview},
  author={Hu, Jiefeng and Shan, Yinghao and Guerrero, Josep M and Ioinovici, Adrian and Chan, Ka Wing and Rodriguez, Jose},
  journal={Renewable and Sustainable Energy Reviews},
  volume={136},
  pages={110422},
  year={2021},
  publisher={Elsevier}
}

@article{nguyen2023distributionally,
  title={Distributionally robust model predictive control for smart electric vehicle charging station with V2G/V2V capability},
  author={Nguyen, Hoang Tien and Choi, Dae-Hyun},
  journal={IEEE Transactions on Smart Grid},
  volume={14},
  number={6},
  pages={4621--4633},
  year={2023},
  publisher={IEEE}
}

@article{micheli2023stochastic,
  title={Stochastic MPC for energy hubs using data driven demand forecasting},
  author={Micheli, Francesco and Behrunani, Varsha and Mehr, Jonas and Heer, Philipp and Lygeros, John},
  journal={IFAC-PapersOnLine},
  volume={56},
  number={2},
  pages={11026--11031},
  year={2023},
  publisher={Elsevier}
}

@misc{seforall2021carbon,
  author       = {{Sustainable Energy for All}},
  title        = {Carbon Emissions Methodology Note},
  year         = {2021},
  month        = {August},
  }

@ARTICLE{gan2021gpfor_detmpc,
  author={Gan, Leong Kit and Zhang, PengFei and Lee, Jaehwa and Osborne, Michael A. and Howey, David A.},
  journal={IEEE Transactions on Sustainable Energy}, 
  title={Data-Driven Energy Management System With Gaussian Process Forecasting and MPC for Interconnected Microgrids}, 
  year={2021},
  volume={12},
  number={1},
  pages={695-704},
  doi={10.1109/TSTE.2020.3017224}
  }

@ARTICLE{babic2023nonparachancempc,  
AUTHOR={Babić, Leon  and Lauricella, Marco  and Ceusters, Glenn  and Biskoping, Matthias },       
TITLE={Data-driven non-parametric chance-constrained model predictive control for microgrids energy management using small data batches},     
JOURNAL={Frontiers in Control Engineering},     
VOLUME={4},
YEAR={2023},
DOI={10.3389/fcteg.2023.1237759},
ISSN={2673-6268}
}

@article{arcos2024mpcecuadorislanded,
title = {Model predictive control-based energy management system for an isolated electro-thermal microgrid in the Amazon region of Ecuador},
author = {Diego Arcos–Aviles and Antonio Salazar and Mauricio Rodriguez and Wilmar Martinez and Francesc Guinjoan},
journal = {Energy Conversion and Management},
volume = {310},
pages = {118479},
year = {2024},
issn = {0196-8904},
doi = {https://doi.org/10.1016/j.enconman.2024.118479},
}

@ARTICLE{hans2020riskaversempcislanded,
  author={Hans, Christian A. and Sopasakis, Pantelis and Raisch, Jörg and Reincke-Collon, Carsten and Patrinos, Panagiotis},
  journal={IEEE Transactions on Control Systems Technology}, 
  title={Risk-Averse Model Predictive Operation Control of Islanded Microgrids}, 
  year={2020},
  volume={28},
  number={6},
  pages={2136-2151},
  doi={10.1109/TCST.2019.2929492}
  }

@article{tavakoli2018cvarresilience,
title = {CVaR-based energy management scheme for optimal resilience and operational cost in commercial building microgrids},
author = {Mehdi Tavakoli and Fatemeh Shokridehaki and Mudathir {Funsho Akorede} and Mousa Marzband and Ionel Vechiu and Edris Pouresmaeil},
journal = {International Journal of Electrical Power \& Energy Systems},
volume = {100},
pages = {1-9},
year = {2018},
issn = {0142-0615},
doi = {https://doi.org/10.1016/j.ijepes.2018.02.022},
}

@article{tobajas2022mpcresilience,
title = {Resilience-oriented schedule of microgrids with hybrid energy storage system using model predictive control},
journal = {Applied Energy},
volume = {306},
pages = {118092},
year = {2022},
issn = {0306-2619},
doi = {https://doi.org/10.1016/j.apenergy.2021.118092},
author = {Javier Tobajas and Felix Garcia-Torres and Pedro Roncero-Sánchez and Javier Vázquez and Ladjel Bellatreche and Emilio Nieto}
}

@article{zandrazavi2022stochasticev,
title = {Stochastic multi-objective optimal energy management of grid-connected unbalanced microgrids with renewable energy generation and plug-in electric vehicles},
journal = {Energy},
volume = {241},
pages = {122884},
year = {2022},
issn = {0360-5442},
doi = {https://doi.org/10.1016/j.energy.2021.122884},
author = {Seyed Farhad Zandrazavi and Cindy Paola Guzman and Alejandra Tabares Pozos and Jairo Quiros-Tortos and John Fredy Franco}
}

@book{modarres2016reliability,
  title     = {Reliability Engineering and Risk Analysis: A Practical Guide},
  author    = {Modarres, Mohammad and Kaminskiy, Mark and Krivtsov, Vasiliy},
  year      = {2016},
  edition   = {3rd},
  publisher = {CRC Press},
  address   = {Boca Raton, FL}
}

@article{houbbadi2019quadratic,
  title={A quadratic programming based optimisation to manage electric bus fleet charging},
  author={Houbbadi, Adnane and Trigui, Rochdi and Pelissier, Serge and Redondo-Iglesias, Eduardo and Bouton, Tanguy},
  journal={International Journal of Electric and Hybrid Vehicles},
  volume={11},
  number={4},
  pages={289--307},
  year={2019},
  publisher={Inderscience Publishers (IEL)}
}

@misc{PURC2025,
  author = {{Public Utilities Regulatory Commission}},
  title  = {{Utility Tariffs (Electricity, Gas, Water) -- Jul. 2025}},
  year   = {2025}
}

@misc{NPA2025,
  author = {{National Petroleum Authority}},
  title  = {{Ex-Pump Prices (Feb. 4, 2025)}},
  year   = {2025}
}

\end{document}